\documentclass[prb,twocolumn]{revtex4}
\usepackage{graphicx,amsmath,amssymb,amsfonts,latexsym,color,dcolumn,bm,mathtools}
\usepackage{verbatim}
\usepackage{epstopdf}
\usepackage[colorlinks=true,linkcolor=blue,citecolor=blue]{hyperref}

\providecommand{\abs}[1]{\left\lvert#1\right\rvert}

\definecolor{blue}{rgb}{0.223,0.223,0.667}
\definecolor{red}{rgb}{0.7,0,0}

\begin{document}

\title{Mechanically mediated microwave frequency conversion in the quantum regime}

\author{F. Lecocq, J. B. Clark,  R. W. Simmonds, J. Aumentado, J. D. Teufel}

\affiliation{National Institute of Standards and Technology, 325 Broadway, Boulder, CO 80305, USA}

\date{\today}

\begin{abstract}
We report the observation of efficient and low-noise frequency conversion between two microwave modes, mediated by the motion of a mechanical resonator subjected to radiation pressure. We achieve coherent conversion of more than $10^{12}~\mathrm{photons/s}$ with a $95\mathrm{\%}$ efficiency and a $14~\mathrm{kHz}$ bandwidth. With less than $10^{-1}~\mathrm{photons \cdot s^{-1}\cdot Hz^{-1}}$ of added noise, this optomechanical frequency converter is suitable for quantum state transduction. We show the ability to operate this converter as a tunable beam splitter, with direct applications for photon routing and communication through complex quantum networks.
\end{abstract}

\maketitle

The interaction between electromagnetic radiation and other quantum systems is ubiquitous in quantum information processing and quantum measurement. Photons can be used to control and measure the quantum state of atoms\cite{Haroche2006}, ions\cite{Leibfried2003}, solid state spins\cite{Togan2010}, superconducting qubits\cite{Blais2004}, or the motion of macroscopic objects\cite{Aspelmeyer2013}. Hence, light fields are ideally suited for coherently connecting nodes in a  quantum networks\cite{Kimble2008}. As in a classical communication network, the ability to shuttle information between different frequency channels is critical in a quantum network. In particular, a frequency converter can be used to route information through complex node architectures or distribute entanglement between systems of vastly different nature and frequency.

While frequency conversion naturally arises in any non-linear system, a frequency converter ideally suited for quantum information processing has to perform a unitary conversion, as signal loss or gain corrupts the quantum signal\cite{Clerk2010}. In the optical domain, using media such as nonlinear crystals\cite{Kumar1992} and optical fibers, it remains challenging to achieve high system efficiency while avoiding or removing unwanted noise generation processes\cite{Srinivasan2012}. In the microwave domain, the development of superconducting mixing elements based on Josephson junctions has enabled near-ideal frequency conversion between microwave signals with appreciable bandwidth, but with very low power handling capability\cite{Abdo2013,Pillet2015}. Additionally, these systems are intrinsically limited to signals in the microwave domain because superconductivity is incompatible with optical light.

\begin{figure}
	\includegraphics[scale=0.97]{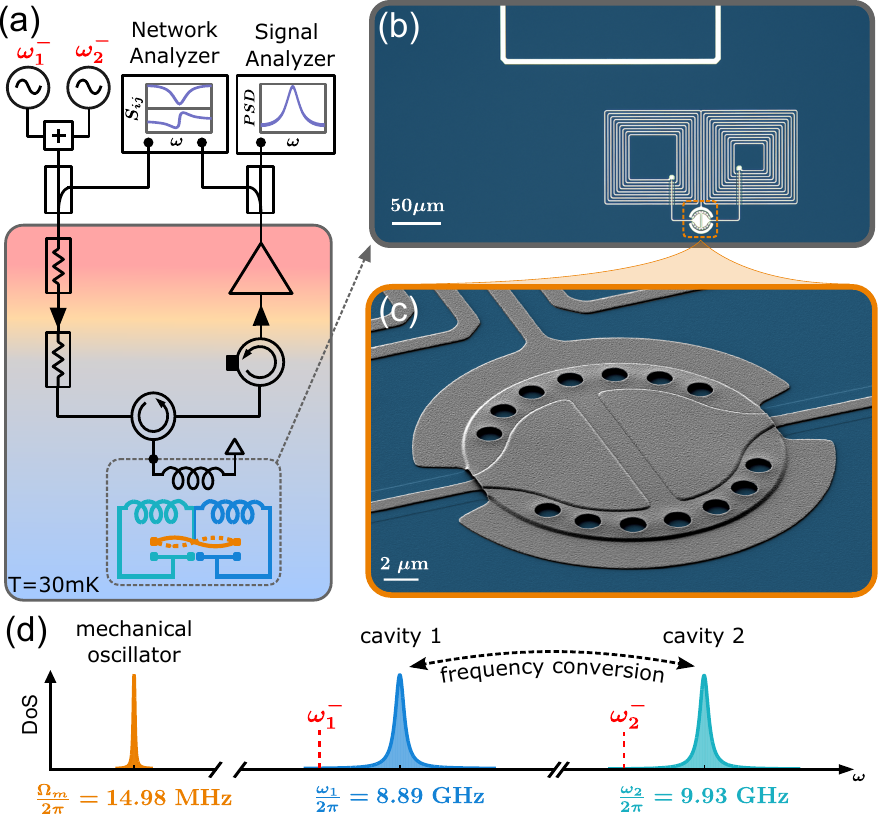}
	\caption{Device description and experimental setup. (a) The device, placed in a cryogenic refrigerator, consists of two inductor-capacitor cavities whose resonance frequencies are tuned by the motion of a mechanically compliant capacitor. Microwave drives are inductively coupled to both cavities via a single port. The reflected signals and the noise emitted by the cavities are amplified and demodulated at room temperature. (b) False-color optical micrograph of the aluminum device on a sapphire substrate (blue). (c) False-color scanning electron micrograph of the mechanically compliant vacuum gap capacitor. (d) Diagram of the density of states (DoS) as a function of frequency. Except for the mechanical linewidth, all the frequencies and linewidths are to scale. The red dashed lines indicate the two parametric drive frequencies, at the lower mechanical sidebands of each cavity, $\omega_{1,2}^{-}=\omega_{1,2}-\Omega_m$.  
    \label{fig1}}
\end{figure}

Recently, experimental breakthroughs in cavity optomechanical systems have enabled the frequency conversion of optical or microwave photons into mechanical phonons\cite{Verhagen2012,Palomaki2013,Lecocq2015}. By coupling a single mechanical element to two different cavities, one can achieve frequency conversion between the two light fields\cite{Safavi-Naeini2011a,Hill2012,Liu2013,Dong2015,Metcalfe2014}. Because of the universal nature of the optomechanical coupling, conversion between microwave and optical frequencies can be achieved\cite{Bochmann2013,Bagci2014,Andrews2014}. However, current implementations have suffered from a low conversion efficiency and from a large added noise due to the residual Brownian motion of the mechanical resonator. Here we report on the observation of parametric frequency conversion between two microwave modes mediated by the motion of a mechanical resonator. We achieve a conversion rate between microwave photons and mechanical phonons that overwhelms the mechanical decoherence rate, ensuring both high conversion efficiency and low added noise.

An optomechanical frequency converter consists of two cavity modes whose resonance frequencies $\omega_1$ and $\omega_2$ are tuned by the position of a single mechanical oscillator of frequency $\Omega_m$. A strong drive is applied at the lower mechanical sideband of each cavity, at the frequencies $\omega_{1,2}^{-}=\omega_{1,2}-\Omega_m$, enabling the coherent exchange of cavity photons and mechanical phonons \cite{Safavi-Naeini2011a}. The full frequency conversion process is as follows: an input microwave probe near the resonance frequency of the first cavity (complex mode amplitude $a_{\mathrm{in}}[\omega_1]$) is down-converted into mechanical motion ($b[\omega_1-\omega_{1}^{-}]=b[\Omega_m]$) then up-converted back to the microwave domain into the output field of the second cavity ($a_{\mathrm{out}}[\Omega_m+\omega_{2}^{-}]=a_{\mathrm{out}}[\omega_2]$), and vice-versa. The performance of the converter can be characterized by its transmission coefficient $t_{1,2}\equiv a_{\mathrm{out}}[\omega_{2,1}]/a_{\mathrm{in}}[\omega_{1,2}]$ and reflection coefficients $r_{1,2}\equiv a_{\mathrm{out}}[\omega_{1,2}]/a_{\mathrm{in}}[\omega_{1,2}]$. As this is a reciprocal process, the transmission is bidirectional and $t_{1}=t_{2}=t$. For an ideal conversion, one needs first to efficiently couple the propagating fields into and out of the cavities; due to finite internal loss of the cavities, the rates at which the intra-cavity fields propagate out of the cavities, $\kappa_{1,2}^{\mathrm{ext}}$, are only a fraction of the total cavity linewidths, $\kappa_{1,2}$, defined as $\eta_{1,2}=\kappa_{1,2}^{\mathrm{ext}}/\kappa_{1,2}$. Second, the photon-to-phonon scattering rates, $\Gamma_{1,2}$, need to overwhelm the mechanical relaxation rate, $\Gamma_{m}$. Their ratios are expressed in terms of cooperativity, $C_{1,2}=\Gamma_{1,2}/\Gamma_{m}$. In the resolved sideband limit, $\kappa_{1,2}\ll\Omega_m$, one obtains $\Gamma_{1,2}=4g_{1,2}^2n_{1,2}/\kappa_{1,2}$, where $g_{1,2}$ are the vacuum optomechanical coupling rates and $n_{1,2}$ are the number of intra-cavity photons induced by each drive. In the weak coupling regime, $\Gamma_{1,2}\ll\kappa_{1,2}$, the scattering parameters take a simple form\cite{Safavi-Naeini2011a}:


\begin{flalign}
	& \left\lvert t\right\rvert^2=\frac{4\eta_{1}\eta_{2}C_1C_2}{\left(1+C_1+C_2\right)^2} \label{T}\\
	& \left\lvert r_{1,2}\right\rvert^2=\left(1-2\eta_{1,2}+2\eta_{1,2}\frac{C_{1,2}}{\left(1+C_1+C_2\right)}\right)^2 \label{R}
\end{flalign}

From Eq. (\ref{T}) one can see that maximum transmission is always obtained for $C_1=C_2$, corresponding to a rate of photon/phonon conversion matched for each cavity. Ideal conversion, $\abs{t}^2=1$, requires high coupling efficiency to each cavity, $\eta_{1,2}=1$ and conversion rates that exceed the mechanical relaxation rate, $C_{1,2}\gg1$. Under these conditions, the circuit is impedance matched and no signal is reflected, $\abs{r_{1,2}}^2=0$. Finally, the bandwidth of the conversion is given by the total full width at half maximum (FWHM) of the mechanical oscillator in presence of the drives, $\Gamma = \Gamma_m + \Gamma_1 + \Gamma_2$.

We realize such an optomechanical frequency converter in the microwave domain using a superconducting circuit in a cryogenic measurement setup, shown in Fig. (\ref{fig1}). The top plate of a vacuum gap capacitor\cite{Cicak2010,Teufel2011} is free to vibrate, and we use the second harmonic mode of motion, resonating at $\Omega_m/2\pi=14.98~\mathrm{MHz}$ with an intrinsic energy relaxation rate $\Gamma_m/2\pi=9.2~\mathrm{Hz}$. The bottom plate of the capacitor is split and each electrode is shunted by a coil inductor, creating two cavity resonances, $\omega_1/2\pi=8.89~\mathrm{GHz}$ and $\omega_2/2\pi=9.93~\mathrm{GHz}$. Both microwave cavities are strongly overcoupled to a single measurement port, setting their linewidths to $\kappa_1/2\pi\approx1.7~\mathrm{MHz}$ and $\kappa_2/2\pi\approx2.1~\mathrm{MHz}$. The internal cavity loss rates are slightly power dependent\cite{Gao2007}, and we calibrate them directly using the cavities' driven responses to obtain $0.93<\eta_{1}<0.96$ and $0.96<\eta_{2}<0.99$. From sideband cooling measurements at multiple cryostat temperatures\cite{Teufel2011b,Weinstein2014}, we calibrate each vacuum optomechanical coupling rate, $g_1/2\pi=145~\mathrm{Hz}$ and $g_2/2\pi=170~\mathrm{Hz}$, as well as the system noise temperature of the measurement setup at each cavity frequency, $T_\mathrm{N}[\omega_{1}]=9.5~\mathrm{K}$ and $T_\mathrm{N}[\omega_{2}]=10.5~\mathrm{K}$. Accordingly, we choose the inductive coupling line as the reference plane for the measurement of the transmission and reflection coefficients. The attenuation and gain of the input and output measurement lines are calibrated\cite{Andrews2014} by measuring the off-resonant reflection coefficient of each cavity as well as the transmission between the cavities in both directions, $t_1$ and $t_2$, in presence of the strong drives.

\begin{figure}
	\includegraphics[scale=1]{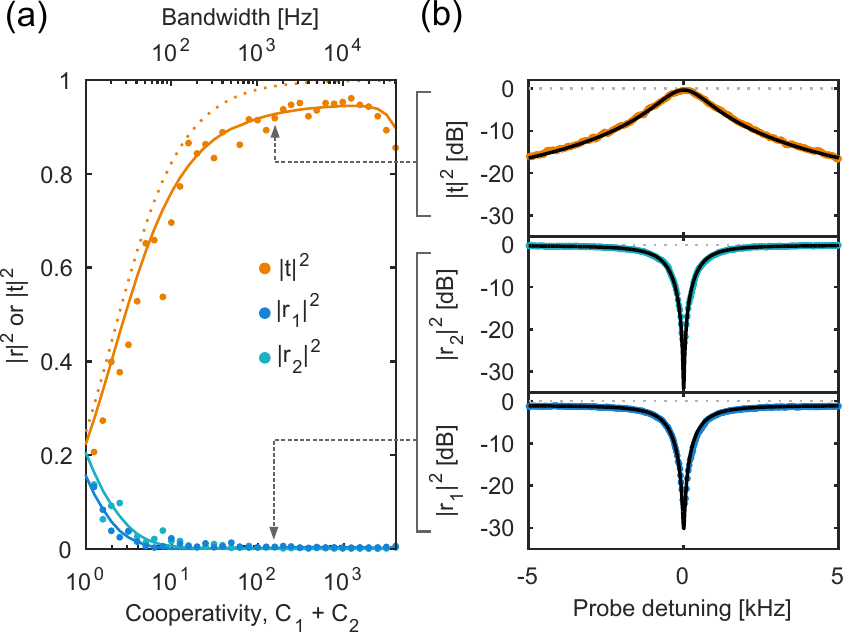}
	\caption{Conversion efficiency. (a) Magnitude of the transmission (orange) and reflection (bright and dark blue) coefficients on resonance, as a function of the total cooperativity $C_1+C_2$. The photon-phonon conversion rate for each cavity is balanced so that $C_1=C_2$. The solid lines are the theoretical predictions from Eqs. (\ref{T}) and (\ref{R}). The dotted line is the inferred internal transmission efficiency, $\abs{t}^2/\eta_{1}\eta_{2}$. The FWHM of the conversion, $\Gamma \approx \Gamma_m(C_1+C_2)$, is shown on the top x-axis. (b) magnitude of the transmission (orange) and reflection (bright and dark blue) coefficients as a function of the probe detuning with respect to the cavity frequencies, for $C_1+C_2=156$, showing a clear Lorentzian shape of width $\Gamma$.
	\label{fig2}}
\end{figure}

The measured scattering parameters are shown in Fig. \ref{fig2}(a) as a function the total cooperativity, $C_1+C_2$, maintaining $C_1=C_2$. As we increase the strength of the drives, the transmission efficiency increases and the reflections decrease, in very good agreement with predictions from Eqs. (\ref{T}) and (\ref{R}) (solid lines). Ultimately, the transmission efficiency is degraded at very high power, due to an increase of the cavity losses. At a total cooperativity of $C_1+C_2=1525$, we measure a transmission of $\abs{t}^2=0.95$, primarily limited by the cavity coupling efficiency $\eta_{1,2}$. The internal conversion efficiency (between intra-cavity modes) is inferred to be almost ideal, $\abs{t}^2/(\eta_{1}\eta_{2})>0.99$. The impedance of the device is well matched, with negligible reflection $\abs{r}^2<0.005$, and the bandwidth reaches $\Gamma/2\pi=14~\mathrm{kHz}$. Finally, we measure a compression of the transmission amplitude by $1~\mathrm{dB}$ when the input power reaches $-75~\mathrm{dBm}$, corresponding to a photon flux of about $5\times10^{12}~\mathrm{photon\cdot s^{-1}}$. This power is more than 2 orders of magnitude larger than that observed in devices based on Josephson junctions\cite{Abdo2013,Abdo2013a}.

\begin{figure}
	\includegraphics[scale=1]{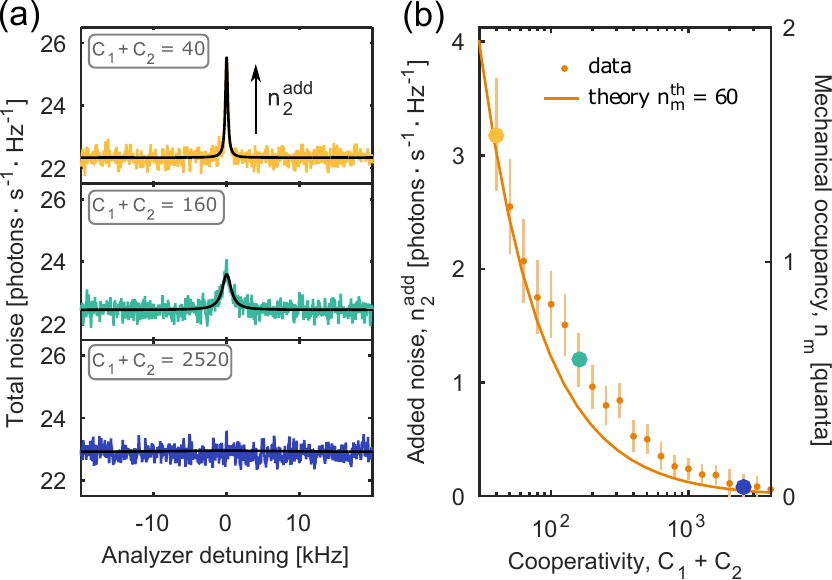}
	\caption{Noise performance. (a) Noise spectrum emitted by cavity 1, around $\omega_{1}=\omega_{1}^{-}+\Omega_m$, for $C_1+C_2=40, 160 ~\mathrm{and}~ 2520$ (with $C_1=C_2$). The measurement noise floor corresponds to a system noise temperature of $T_\mathrm{N}[\omega_{1}]=9.5~\mathrm{K}$. The noise added by the frequency converter appears as a Lorentzian of width $\Gamma$ and peak amplitude $n_{2}^{\mathrm{add}}$. (b) Noise added by the frequency conversion as a function of the total cooperativity $C_1+C_2$. The solid line is the theoretical prediction for an equilibrium mechanical thermal occupancy of $n_m^{\mathrm{th}} = 60$. The right y-axis is the corresponding mechanical occupancy, following Eq. (\ref{Nadd}). The yellow, green and blue dots correspond to the spectra shown in (a). 
	\label{fig3}}
\end{figure}

In addition to being efficient, an ideal frequency converter should not add noise. In an optomechanical frequency
converter, any residual thermal motion of the mechanical oscillator will add noise during the conversion process. This puts more stringent bounds on the scattering rates, which now need to overwhelm the thermal decoherence rate such that $\Gamma_{1,2}\gg \Gamma_mn_m^{\mathrm{th}}$, where $n_m^{\mathrm{th}}$ is the equilibrium mechanical thermal occupancy. The added noise is generally characterized by the effective number of noise photons added to an input signal: $S_{\mathrm{out}}[\omega_{2,1}]=\abs{t}^2\left(S_{\mathrm{in}}[\omega_{1,2}]+n_{1,2}^{\mathrm{add}}\right)$, where $S_{\mathrm{in}}[\omega_{1,2}]$ and $S_{\mathrm{out}}[\omega_{2,1}]$ are the quantum noise spectra\cite{Clerk2010} at the input of one cavity and at the output of the other cavity, respectively. The added noise $n_{1,2}^{\mathrm{add}}$ is: 
\begin{equation}
	n_{1,2}^{\mathrm{add}}=\frac{n_m^{\mathrm{th}}}{\eta_{1,2}C_{1,2}}>2n_m \label{Nadd}
\end{equation}
where $n_m=n_m^{\mathrm{th}}/(1+C_1+C_2)$ is the final mechanical occupancy in presence of the optomechanical drives.

Typical measured noise spectra at the output of cavity 1 are shown in Fig. \ref{fig3}(a) for three drive strengths ($C_1+C_2=40, 160 ~\mathrm{and}~ 2520$, with $C_1=C_2$). The noise added by the mechanical motion appears as a Lorentzian of width $\Gamma$ and peak amplitude $n_{2}^{\mathrm{add}}$. In Fig. \ref{fig3}(b), we show the added noise and the corresponding mechanical occupancy, as a function the total cooperativity, $C_1+C_2$. As expected, the added noise decreases with cooperativity and, at the highest cooperativity, the added noise is negligible compared to the noise floor, with $n_{2}^{\mathrm{add}}<0.1$. The data are in good agreement with predictions from Eq. (\ref{Nadd}) for an effective bath temperature $n_m^{\mathrm{th}}=60$, slightly warmer than the expected value at the cryostat base temperature, $T=30~\mathrm{mK}$. We point out that spectra measured over a wider frequency span show no evidence of internal cavity thermal occupancy at these drive strengths \cite{Weinstein2014}. At the highest drive powers, saturation of the measurement setup eventually raises the noise floor (see Fig. \ref{fig3}(a)). We emphasize that with an added noise much smaller than a single photon and a high conversion efficiency, we have demonstrated the first realization of an optomechanical frequency converter in the quantum regime.

\begin{figure}
	\includegraphics[scale=1]{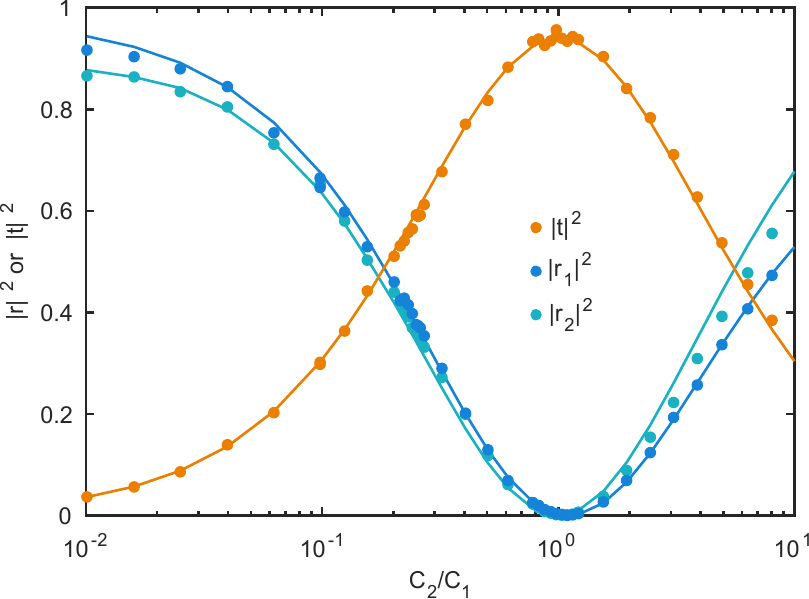}
	\caption{Tunable beam splitter. The magnitude of the transmission (orange) and reflection (bright and dark blue) coefficients on resonance, as a function of the ratio $C_2/C_1$ for a fixed $C_1=400$. The solid lines are the theoretical predictions from Eqs. (\ref{T}) and (\ref{R}). The device can be tuned from a near-ideal mirror ($C_2/C_1\ll1$) to fully transparent ($C_2=C_1$), and for $C_2/C_1\approx 0.2$ one realizes a 50/50 beam splitter.
		\label{fig4}}
\end{figure}

A direct application of our device is the coherent routing of photons, controlled by the strength of the drives. In the regime of high cavity coupling, $\eta_{1,2}\approx1$, and high conversion rates, $C_{1,2}\gg1$, the device can operate as a tunable beam splitter. Its transparency is simply tuned by the ratio of the drive strengths [see Eqs. (\ref{T}) and (\ref{R})]. In Fig. \ref{fig4} we show the measured scattering parameters as a function the cooperativity ratio, $C_2/C_1$. We fix $C_1=400$ to be in the quantum regime $C_1+C_2 \gg n_m^{\mathrm{th}}$ and vary the cooperativity $C_2$. At low cooperativity ratio, both cavities reflects the incoming signal with very little losses. When the cooperativities are matched, $C_1=C_2$, we retrieve the near-ideal transmission shown in Fig. \ref{fig2}. For $C_2/C_1\approx3\pm2\sqrt{2}$, one obtains $\abs{t}^2\approx\abs{r_{1,2}}^2\approx0.5$, and the device acts as a 50/50 beam splitter. This would allow for the creation of quantum state superposition between the propagating fields. Furthermore, the transparency of the beam splitter can be tuned in-situ, in a sub-microsecond timescale, by changing the drives strengths.

Looking forward, increasing the conversion bandwidth by an order of magnitude would make it directly compatible with state of the art microwave single photon sources\cite{Kindel2015}. While such a bandwidth is achievable in our system by simply increasing the drive strengths, it is at the expense of an increase in cavity loss. This could be mitigated by an improvement of the  vacuum optomechanical coupling rates, $g_{1,2}$, or by an optimization of the cavities widths, $\kappa_{1,2}$. Ultimately the bandwidth is limited to the mechanical frequency $\Omega_m$, and one need to maintain $\kappa_{1,2}<\Omega_m$ to avoid unwanted parametric gain and added noise\cite{NoteCRterm}. Finally, our results show that mechanical resonator can act as ideal mixing element. The implementation of more complex mode structure and pumping scheme would allow for non-reciprocal transmission or amplification\cite{Metelmann2015}.

Contribution of the U.S. government, not subject to copyright.

\end{document}